\documentclass{sig-alternate}

\usepackage{subfigure}
\usepackage{rotating}
\usepackage{multirow}
\usepackage{color}

\usepackage[english]{babel}

\usepackage{url}
\usepackage{amssymb}
\usepackage{amsmath}\usepackage{amssymb,stmaryrd}

\newcommand{\ra}[1]{\renewcommand{\arraystretch}{#1}}

\newcommand{\PRISM}{\textsc{PRISM} }

\ifx\property\undefined
\newtheorem{property}{Property}
\fi

\newcommand{\Probs}{\mathtt{P}}

\newcommand{\Probr}{\mathtt{R}}

\newcommand{\Path}{\mathit{Path}}
\newcommand{\D}{\mathcal{D}}

\newcommand{\probmeasure}{\operatorname{\Pr}}

\newcommand{\TF}{{\sf F}}
\newcommand{\TU}{{\sf U}}

\newcommand{\TX}{{\sf X}}
\newcommand{\TC}{{\sf C}}
\newcommand{\TP}{{\sf P}}
\newcommand{\TR}{{\sf R}}

\newcommand{\statelabel}[1]{\textsf{#1}}
\newcommand{\TermsAndConditions}{\statelabel{TermsAndConditions}}
\newcommand{\Main}{\statelabel{Main}}
\newcommand{\TopApps}{\statelabel{TopApps}}
\newcommand{\LastSevenDays}{\statelabel{Last7Days}}
\newcommand{\PeriodSelector}{\statelabel{PeriodSelector}}
\newcommand{\AppsInPeriod}{\statelabel{AppsInPeriod}}
\newcommand{\Settings}{\statelabel{Settings}}
\newcommand{\UseStop}{\statelabel{UseStop}}
\newcommand{\Stats}{\statelabel{Stats}}
\newcommand{\UsageBarChartTopApps}{\statelabel{UsageBarChartTopApps}}
\newcommand{\UsageBarChartStats}{\statelabel{UsageBarChartStats}}
\newcommand{\Feedback}{\statelabel{Feedback}}
\newcommand{\UsageBarChartAppsInPeriod}{\statelabel{UsageBarChartApps}}
\newcommand{\Info}{\statelabel{Info}}
\newcommand{\Task}{\statelabel{Task}}

\newcommand{\PropFour}[2]{\mathit{probToReach}(#1)\mathit{from}(#2)\mathit{DuringOneSession}}
\newcommand{\PropFive}[2]{\mathit{stepsToReach}(#1)\mathit{from}(#2)}

\newcommand{\eqdef}{\mbox{\small$\stackrel{\mbox{\tiny$\triangle$}}{=}$}}

\newcommand{\OverallAP}{\textsf{Overall Viewing}}
\newcommand{\TimePartitionedAP}{\textsf{Time-partitioned Viewing}}

\begin{document}


\title{Probabilistic Formal Analysis of App Usage\\ to Inform Redesign}

\numberofauthors{1} 
\author{
\alignauthor
Oana Andrei, Muffy Calder, Matthew Chalmers, Alistair Morrison, 
Mattias Rost \\
\affaddr{School of Computing Science}\\
       \affaddr{University of Glasgow}\\
       \affaddr{Glasgow, UK}\\
       \email{firstname.lastname@glasgow.ac.uk}
}

\maketitle

\begin{abstract}
  This paper sets out a process of app analysis intended to support
  understanding of use but also redesign.  From usage logs we infer
  activity patterns -- Markov models -- and employ probabilistic
  formal analysis to ask questions about the use of the app.  The core
  of this paper's contribution is a bridging of stochastic and formal
  modelling, but we also describe the work to make that analytic core
  utile within a design team.  We illustrate our work via a case study
  of a mobile app presenting analytic findings and discussing how they
  are feeding into redesign. We had posited that two activity patterns
  indicated two separable sets of users, each of which might benefit
  from a differently tailored app version, but our subsequent analysis
  detailed users' interleaving of activity patterns over time --
  evidence speaking more in favour of redesign that supports each
  pattern in an integrated way. We uncover patterns consisting of
  brief glances at particular data and recommend them as possible
  candidates for new design work on widget extensions: small displays
  available while users use other apps.

\end{abstract}




\section{Introduction}

Good design of user-intensive applications (henceforth referred to as
{\em apps}) is challenging -- because users are seldom homogeneous or
predictable in the ways they navigate around and use the functionality
presented to them.  Different populations of users will engage in
different ways, and redesign may be desirable or even required to
support populations' different styles of use.

This simple hypothesis raises many questions, including: how should we
identify the different populations, what characterises a population,
and does that characterisation evolve, e.g. over an individual use
session, and/or over a number of sessions over days and months?  This
paper attempts to answer these questions, in the context of informing
future redesign of an app.  We propose that formal, probabilistic
analysis of {\em inferred} patterns of {\em logged} app usage is key,
and we refer to these patterns as {\em activity patterns}.  Our
concept of population is therefore based on inferred temporal
behaviours, i.e. activity patterns, rather than on static or slowly
changing user attributes such as gender and age.  The novelty of our
approach is realising the concept of population through a combination
of three powerful `ingredients': 
\begin{itemize}
\item {\em inference} of Markov models of activity patterns from
  automatically logged data on user sessions,
\item characterisation of the activity patterns by probabilistic {\em
    temporal logic} properties using model checking,
\item {\em longitudinal} analysis of usage data drawn from different
  time cuts (e.g. logged sessions over the first day, first month,
  second month, etc.).
\end{itemize}

The focus of this paper is how populations of users are identified and
characterised, and how they evolve.  The contribution is defining the
whole process from identifying questions that give us insight into an
application, to event and attribute logging, data pre-processing and
abstraction from logs, model inference, temporal logic property
formulation using the probabilistic temporal logic PCTL with
rewards~\cite{BaierKatoen-MCbook}, and visualisation of results.  We
illustrate throughout with a case study of
AppTracker~\cite{BellCFHMRRS13}, a freely available mobile app that
allows users to collect quantitative statistics about the usage of
apps installed on one's phone.  AppTracker users can measure, for
example, which apps one uses the most, how much time one has spent on
each app, the average daily use, and the most and least active
days. The application comprises several screens that display
statistics in numerical as well as graphical formats.

A notable conclusion of our work is that, while our analysis of
AppTracker's use identifies several clearly distinct activity
patterns, it also reveals the distribution of activity patterns over
the population of users and over time. For AppTracker, this mitigates
against a simple partitioning of the app into two different versions,
each specific to one activity pattern. However, our analysis does
offer a more principled way of selecting glanceable information.

The paper is organised as follows.  First we give an overview of our
approach followed by a technical background including probabilistic
models, logics, and inference. Section~\ref{sect:casestudy} contains
an overview of AppTracker and Sect.~\ref{sect:methods} describes how
the methods of our approach are applied. Section~\ref{sect:analysis}
presents results from the analysis and Sect.~\ref{sect:redesign}
discuss how they offer insights into redesign of the app. Related work
is discussed in Sect.~\ref{sect:relatedwork}, followed by a
discussion about generalising our approach to analysis of other type
of apps. Conclusions and future work are in Sect.~\ref{sect:conclusions}.

\section{Approach}\label{sect:approach}

Our approach begins with {\em instrumenting} an app, so that we can
log usage behaviours.  In our case, the app was instrumented by the
developers, using the SGLog data logging
infrastructure~\cite{Hall:2009}. SGLog detects user {\em events} (such
as button taps or screen changes within the app), stores log entries
in a local text file on the device, and periodically uploads this data
back to the developers' servers. The raw logged data are processed to
create sets of user traces expressed in terms of higher level actions.
The higher level actions are carefully selected to relate to the
intended analysis, namely to the underlying {\em atomic propositions}.
The choice of these propositions is key: they determine the scope of
properties that can be revealed by temporal property analysis and also
determine the dimensions of the state space underlying the model. They
are defined jointly by analysts and developers; in the case study, the
propositions are the high level states relating to the core
functionality of app, e.g. {\em select main view}, {\em show selected
  statistics of device and app use}, {\em close app}, etc. The sets of
traces can be partitioned into different time cuts, e.g. the first $7$
days of usage, the first week, the second month, so that we can
determine how activity patterns evolve over time and experience with
the app.
 
We run a non-linear optimisation algorithm for parameter estimation on
a set of user traces in order to infer admixture models of {\em
  activity patterns}. The activity patterns represent the set of user
traces {\em bottom up} rather than imposing a set of categorising features
{\em a priori}.  Each activity pattern is a discrete-time Markov
chain, and we can then characterise each user trace as a weighted mix
of activity patterns.  We note an aspect of the paper's contribution
here: to the best of our knowledge, inferring such temporal structures
has not been described in prior work outside our group.

We then {\em hypothesise} temporal probabilistic properties, expressed
in the probabilistic temporal logic PCTL extended with
rewards~\cite{BaierKatoen-MCbook,KwiatkowskaNP07}, to explore the
inferred activity patterns.  A typical probabilistic temporal property
is {\em the expected number of visits to a given state within $N$
  steps from the start of a session, for each activity pattern and for
  different time cuts.} Analysts define the temporal properties for
various admixture models and time cuts, and discuss the results with
developers. The discussions prompt the analysis of more properties,
models and time cuts, and hypotheses to test, but more generally
provide new insights into app usage and afford new ideas for redesign
that are solidly grounded in observed activity patterns.

This is unlike standard use of model checking, where we are given a
discrete-time Markov chain with fixed rates and requirements as
temporal properties (which we attempt to verify).  Our approach here
is a combination of bottom up statistical inference from user trace
data and top down probabilistic temporal logic analysis of the
inferred models.

\section{Technical Background}
\label{sect:background}

We assume familiarity with Markov models, bigram models and
Expectation-Maximisation 
algorithms~\cite{Demp1977,Murphy02dynamicbayesian,StollerBSGHSZ11},
the probabilistic temporal logic PCTL and probabilistic model checking
for DTMC~\cite{BaierKatoen-MCbook,KwiatkowskaNP07}; basic definitions
are below.

\subsection{Discrete-time Markov chains}

A {\em discrete-time Markov chain} (DTMC) is a tuple
$\D=(S,\bar{s},\Probs,\ell)$ where: 
\begin{itemize}
\item $S$ is a set of states;
\item $\bar{s}\in S$ is the initial state; 
\item $\Probs:S\times S\rightarrow
[0,1]$ is the transition probability function (or matrix) such that
for all states $s\in S$ we have $\sum_{s'\in S}\Probs(s,s')=1$; 
\item $\ell:S\rightarrow 2^\mathcal{A}$ is a labelling function
  associating to each state $s$ in $S$ a set of valid atomic
  propositions from a set $\mathcal{A}$. 
\end{itemize}
A {\em path} (or execution) of a DTMC is a non-empty sequence
$s_0s_1s_2\ldots$ where $s_i\in S$ and $\Probs(s_i,s_{i+1})>0$ for all
$i\geq 0$.  A transition is also called a {\em time-step}.


\subsection{Probabilistic logics and model checking}

{\em Probabilistic Computation Tree Logic} (PCTL)~\cite{BaierKatoen-MCbook}
 allows one to express a probability measure of
the satisfaction of a temporal property.  The syntax is:
\vspace{-0.1cm}
\begin{center}
  \begin{tabular}{rl} 
    {\em State formulae} & $\phi ::=
    \mathit{true} \mid a \mid \lnot \phi \mid \phi \land \phi \mid
    \Probs_{\bowtie\, p}[\psi]$
    \\
    {\em Path formulae} & $ \psi ::= \TX\, \phi \mid \phi\,
    \TU^{\leq n}\, \phi$
\end{tabular}
\end{center}
\vspace{-0.1cm} where $a$ ranges over a set of atomic propositions
$\mathcal{A}$, $\bowtie\,\in \{\leq, <, \geq, >\}$, $p\in [0,1]$, and
$n\in \mathbb{N}\cup \{\infty\}$. State formulae are also called {\em
  temporal properties}.

A state $s$ in a DTMC $\D$ satisfies an atomic proposition $a$ if
$a\in \ell(s)$.  A state $s$ satisfies a state formula
$\Probs_{\bowtie\, p}[\psi]$, written $s\models \Probs_{\bowtie\,
  p}[\psi]$, if the probability of taking a path starting from $s$ and
satisfying $\psi$ meets the bound $\bowtie\, p$, i.e.,
$\probmeasure_s\{\omega\in \Path^\D(s) \mid \omega\models \psi\}
\bowtie\, p$, where $\probmeasure_s$ is the probability measure
defined over paths from state $s$.  The path formula $\TX\, \phi$ is
true on a path starting with $s$ if $\phi$ is satisfied in the state
following $s$; $\phi_1\, \TU^{\leq n}\, \phi_2$ is true on a path if
$\phi_2$ holds in the state at some time step $i\leq n$ and at all
preceding states $\phi_1$ holds. The propositional operators
$\mathit{false}$, disjunction and implication can be derived using
basic logical equivalence. In this paper we use the \emph{eventually}
operator $\TF$ where $\TF^{\leq n}\,\phi\equiv \mathit{true}\,
\TU^{\leq n}\,\phi$.  If $n =\infty$ then superscripts are omitted.

\subsection{The probabilistic model checker \PRISM}

The \PRISM probabilistic model checker~\cite{KwiatkowskaNP11} allows
us to leave the bound $\bowtie\ p$ unspecified and computes the
satisfaction probability by verifying the property
$\Probs_{=?}\left[\,\psi\,\right]$.  Additionally, \PRISM allows for
{\em experimentation}: the verification of an open formula, when the
range, and step size of the variable(s) are specified.  \PRISM
supports a {\em reward}-based extension of PCTL called {\em rPCTL}. A
reward structure assigns non-negative real values to states and/or
transitions. 
We employ rewards assigned to transitions and \emph{cumulative} and
\emph{reachability} reward properties. Then whenever a transition is
taken, the reward associated with it is earned. The cumulative reward
property, $\Probr_{=?}  \left[\, \TC^{\leq N}\right]$, computes the
accumulated reward along {\em all} paths within $N$ time-steps. The
reachability reward property, $\Probr_{=?} \left[\, \TF\ \phi
  \,\right]$, computes the reward accumulated along {\em all} paths
until the state formula $\phi$ is satisfied. The reward properties are
usually annotated with specific reward structures.
 
The PRISM default is to reason over state formulae from the initial
state of the DTMC under analysis. Filtered probabilities check for
properties that hold {\em from sets of states} satisfying given
propositions.  In the examples illustrated in this paper we always use
$\mathtt{state}$ as the filter operator: e.g.,
$\mathtt{filter}(\mathtt{state}, \phi, \mathit{condition})$ where
$\phi$ is a state formula and $\mathit{condition}$ a Boolean
proposition uniquely identifying a state in the DTMC.

\begin{figure*}[!t]
  \centering 
  \subfigure[Main menu]{\includegraphics[scale=0.128]{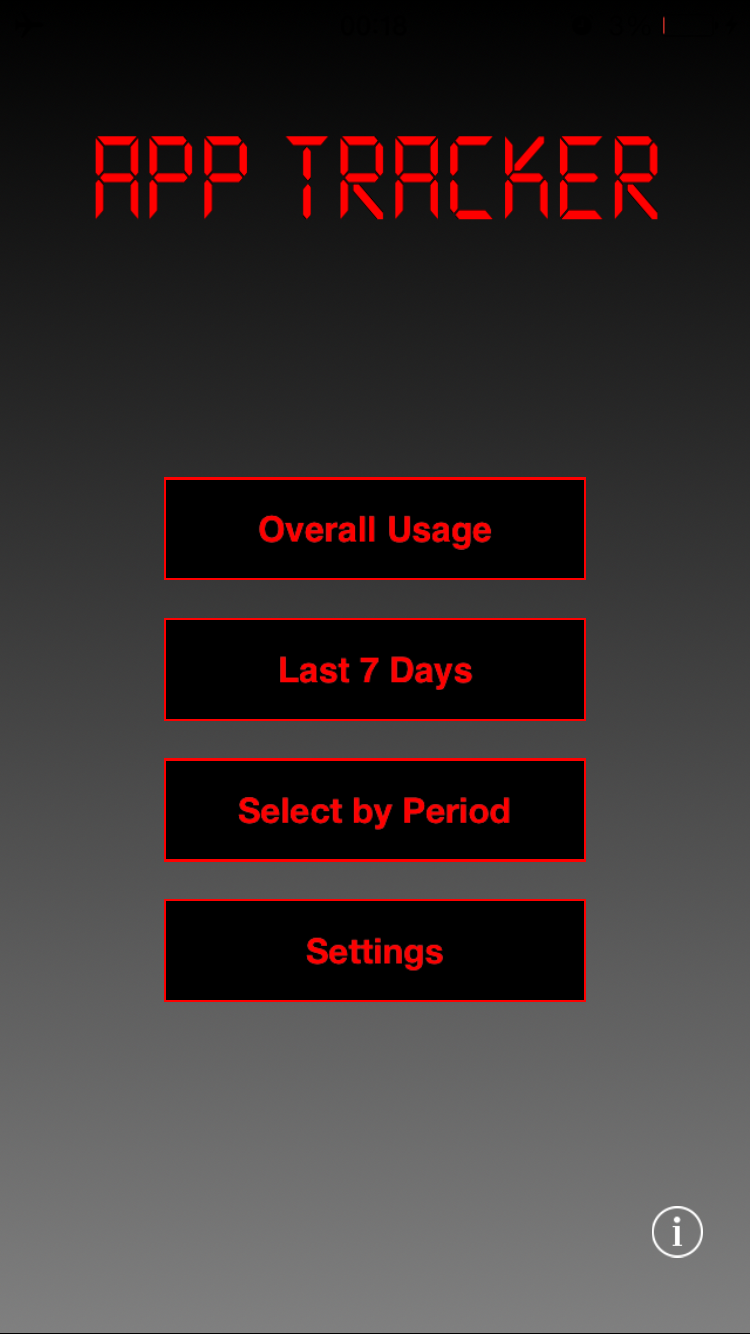}\label{fig:menu}}
  \quad
  \subfigure[Overall stats]{\includegraphics[scale=0.15]{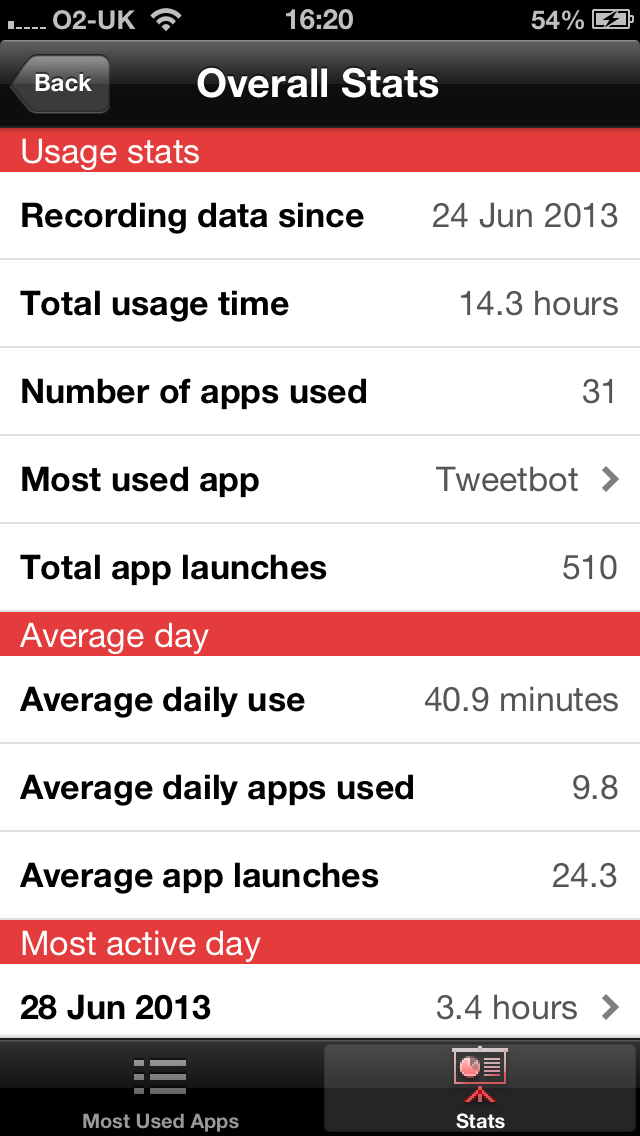}\label{fig:stats}}
  \quad  
  \subfigure[Device usage for one day]{\includegraphics[scale=0.15]{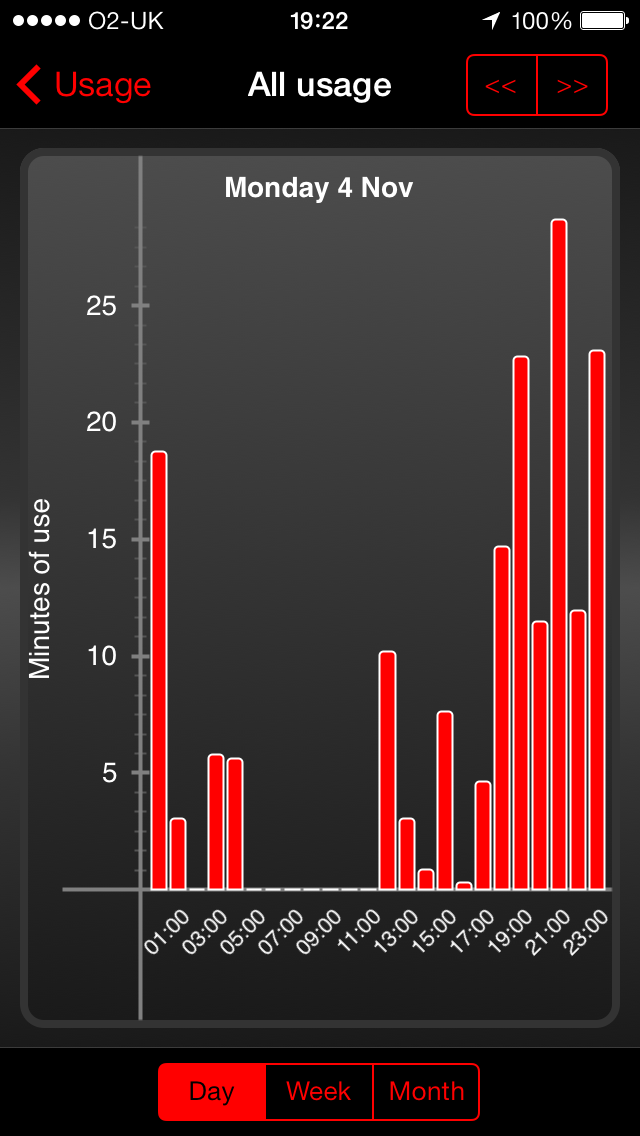}\label{fig:day}} 
  \caption{Screenshots from AppTracker}
\label{fig:screenshots}
\end{figure*}

\subsection{Inference of admixture bigram models} 

Given a vocabulary $V$ of size $n$, an observation sequence (or trace)
over $V$ is a finite non-empty sequence of symbols from $V$. Let $x$
be a data sample of $M$ traces over $V$, $x=\{x_1,\ldots, x_M\}$.
Consider $K$ $n\times n$ transition matrices denoted $\Phi_k$, for
$k=1,\ldots, K$, such that $\Phi_{kij}$ denotes the probability of
moving to state $j$ from $i$.  At any point in time $\Phi_k$ is used
by the trace $x_m$ with probability $\Theta_{mk}$.  Let
$\lambda=\{\Phi_k,\Theta_{mk} \mid k=1,\ldots, K, \ m=1,\ldots, M\}$
be the parameters of the statistical model. Then we estimate the
likelihood of observing the trace $x_m$ given $\lambda$, i.e.,
$p(x_m|\lambda)$, which takes the form:
$$
p(x_m|\lambda) = \prod_{i=1}^{n}\prod_{j=1}^n\Big(\sum_{k=1}^K\Theta_{mk}\Phi_{kij}\Big)^{x_{mij}}
$$
We use a non-linear optimisation algorithm to find the (locally)
maximum likelihood parameter estimation. Our algorithm of choice is
the Expectation-Maximisation (EM) algorithm~\cite{Demp1977} and we
restart it whenever the log-likelihood $\ln(p(x_m)|\lambda)$ has
multiple-local maxima. Each trace $x_m$ can be represented as a DTMC
as follows:
\begin{itemize}
\item the set of states $S=\{0,1,\ldots,|V|\}$ where $|V|$ is the size of
$V$, 
\item the initial state is $0$,
\item the transition probability matrix is
the $n\times n$ transition-occurrence matrix such that $x_{mij}$ on
position $(i,j)$ gives the number of times the pair $(x_{mi},x_{mj})$
occurs in the trace $x_m$, 
\item $\ell:S\rightarrow V$ is a bijective
function such that $\ell(0)$ is the symbol on the first position in
the trace $x_m$. 
\end{itemize}
Let all traces in $x$ start with the same symbol. Then the EM
algorithm finds maximum log-likelihood estimates for $\lambda$
consisting of $K$ DTMCs with the same sets of states $S$ and initial
state $0$, and an $M\times K$ weighting matrix $\Theta$. The result is
a $\Theta$-weighted mixture of the $K$ DTMCs $\Phi_k$ forming an {\em
  admixture bigram model}: $\Theta_{mk}$ indicates the probability of
using $\Phi_k$ to transition between states. The model is bigram
because only dependencies between adjacent symbols in the trace are
considered.

\section{Case study: AppTracker}
\label{sect:casestudy}


AppTracker is an iOS application that provides a user with information
on the usage of his/her device. The app operates on iPhones, iPads and
iPod Touches, running in the background and monitoring the opening and
closing of apps as well as the locking and unlocking of the
device. AppTracker was released in August 2013. To date it has been
downloaded over 35,000 times.

AppTracker's interface displays a series of charts and statistics to
give insight into how long one is spending on one's device, the most
used apps, and how these stats fluctuate over
time. Figure~\ref{fig:screenshots} shows three views from the app. The
AppTracker interface has a main menu screen (\Main), presenting four
main options (Fig.~\ref{fig:menu}).  The first menu item, {\em Overall
  Usage}, contains quick summaries of all the data recorded since
AppTracker was installed: the view \TopApps\ shows a list of the
user's most-used apps and the view \Stats\ shows summary statistics
such as the number of apps used in an average day
(Fig.~\ref{fig:stats}).  The second menu item, {\em Last 7 Days},
displays a chart limited to the last week's activity, showing a
stacked bar graph of usage of the top 5 apps during that period; this
view is called \LastSevenDays. The third menu item, {\em Select by
  Period}, opens up the \PeriodSelector\ view where a user can see
usage statistics by any day, week or month, and drill down to a
particular period of interest. For example, one could investigate
which apps one used the most last Saturday, see how time one spent on
Facebook varied each day across last month, or examine patterns of use
over a particular day (Fig.~\ref{fig:day}). The final menu option,
{\em Settings}, allows a user to start and stop the tracker, or to
reset his/her recorded data.

\begin{figure*}[!t]
  \centering 
  \includegraphics[scale=0.45]{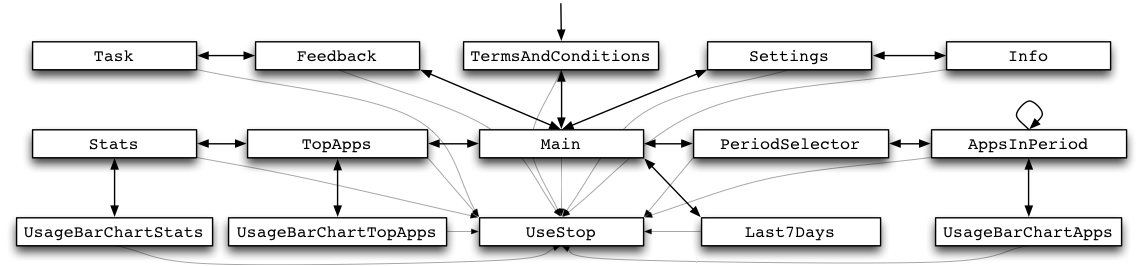}
  \caption{AppTracker state diagram}
\label{fig:diagram}
\end{figure*}

A {\em Terms and Conditions} screen is shown to a user on first
launch, that explains the nature of AppTracker as a research project,
describes all the data that will be recorded during its use and
provides contact details to allow the user to opt out of the study at
any time. These terms must be agreed to before the user has access to
any other part of the app. The overall functional behaviour of
AppTracker is illustrated in Figure~\ref{fig:diagram}.
 

\begin{table}[t!]
\ra{1.2}
\caption{The AppTracker higher level states}
{\scriptsize
\begin{center}
\begin{tabular}{|r|r|l|}
  \hline 
  \textbf{Id} & \textbf{State} &  \textbf{Description} \\
  \hline
  0 & \TermsAndConditions  & Terms and conditions page\\
  \hline
  1& \Main  & Main display \\
  \hline
  2 & \TopApps  & Summary of all recorded data\\
  \hline
  3 & \LastSevenDays  & The last 7 days of top 5 apps used\\
  \hline
  4 & \PeriodSelector  & Choose a time period \\
& & to see app usage\\
  \hline
  5 & \AppsInPeriod  & Apps used for a selected period\\
  \hline
  6 & \Settings  & Settings view\\
  \hline
  7 & \UseStop  & Close AppTracker\\
  \hline
  8 & \Stats  & Select statistics of app usage\\
  \hline
  9 & \UsageBarChartTopApps  & App usage when picked from \TopApps\\
  \hline
  10 & \UsageBarChartStats  & App usage  when picked from \Stats\\
  \hline
  11 & \Feedback  & Screen for giving feedback\\
  \hline
  12 & \UsageBarChartAppsInPeriod  & App usage when picked from \\
 & &   \AppsInPeriod\\
  \hline
  13 & \Info  & Information about the app\\
  \hline
  14 & \Task  & A feedback question chosen from \\
& & the \Feedback\ view\\
  \hline
\end{tabular}
\end{center}
}
\label{table:states}
\vspace{-3.5mm}
\end{table}

\section{Methods Applied to AppTracker}
\label{sect:methods}

\subsection{Preparing the raw SGLog data}

\paragraph{SGLog data}  Data collected by SGLog~\cite{Hall:2009} (the
logging framework used to instrument AppTracker) consists of
timestamped logs of events, such as user actions. Each log contains
information about the user and device, and the event that took place.
For our analysis, we are interested in the events resulting in a
switch between views within the app, and from that we focus on which
view the user transitions to.  We therefore transform the raw logs of
events into user traces of views, which is a list of views visited
within the app. A special view denotes when the user leaves the app
(\UseStop). This results in a total of 15 unique views listed in
Table~\ref{table:states}.  These views and transitions between views
are used as states and transitions between states, and relate directly
to the underlying atomic propositions. Figure~\ref{fig:usertrace}
illustrates a fragment of a logged user trace: information about the
user's device, start and end data of AppTracker usage, and the first
session. The log data is stored in a MySQL database by the SGLog
framework. Raw data is extracted from the database and processed using
JavaScript to obtain user traces.

\begin{figure*}[!t]
\centering
{\scriptsize
\begin{verbatim}
[{"deviceid":"xx:xx:xx:xx:xx:xx","totalevents":230,"firstSeen":"2013-08-20 09:10:59","lastSeen":"2014-03-24 09:57:32", 
  "sessions":[[{"timestamp":"2013-08-20 09:11:02","data":"TermsAndConditions"},{"timestamp":"2013-08-20 09:11:23", "data":"Main"},
  {"timestamp":"2013-08-20 09:11:46","data":"TopApps"},{"timestamp":"2013-08-20 09:11:50", "data":"Main"},{"timestamp":"2013-08-20 
  09:11:52","data":"Last7Days"},{"timestamp":"2013-08-20 09:11:56", "data":"Main"},{"timestamp":"2013-08-20 :11:59",   
  "data":"PeriodSelector"},{"timestamp":"2013-08-20 09:12:04", "data":"Main"},{"timestamp":"2013-08-20 09:12:06","data":"UseStop"}],...
\end{verbatim}
}
\vspace{-.24cm}
\caption{Example of a user trace fragment.}
\label{fig:usertrace}
\end{figure*}

All data analysed for this paper was gathered between August 2013 and
May 2014, from 489 users. The average time spent within the app per
user is 626s (median 293s), the average number times going into the
app is 10.7 (median 7), the average user trace length is 73.6 view
transitions.  (median 46).

\paragraph{Time cuts} We use JavaScript to segment the log data into
time cuts of the interval form $[d_1,d_2)$, which returns the user
traces from the $d_1$-th up until the $d_2$-th day of usage.

\paragraph{Computing transition-occurrence matrices}  We use JavaScript for
mapping each user trace to a $15\times 15$ transition-occurrence
matrix: the number in position $(i,j)$ denotes how many times the
$i$-th view is followed by the $j$-th view in the trace. A transition
represents an action performed by the user to switch to one of the 15
views obtained from the SGlog data, including \UseStop.

\subsection{Inferring activity patterns}
\label{sect:inference}

For each value of $K$ and time cut of the app usage we obtain $K$
DTMCs with $15\times 15$ transition matrices called {\em activity
patterns} and an $M\times K$ matrix $\Theta$, where $M$ is the number
of user traces, and with each row a distribution over the $K$ activity
patterns. 

For each activity pattern $\mathit{APk}$, for $k=1,\ldots,K$, we
generate automatically a \PRISM model with one variable for the views
of the app with values ranging from 0 to 14.  For each state value of
$x$ we have a \PRISM command defining all possible 15 probabilistic
transitions. Each value $\Phi_{kij}$ is the transition probability
from state $x=i$ to the updated state $x'=j$ in activity pattern
$APk$, for all $i,j=0,\ldots,14$. For each state value we associate
the label corresponding to a higher level state in AppTracker (see the
mapping in Table~\ref{table:states}) as well as a reward structure
which assigns a reward of 1 to that state. The \PRISM file for each
activity pattern also includes a reward structure assigning a rewards
of 1 to each transition (or time step) in the DTMC. All \PRISM models
have at most 15 states and at most 51 transitions for all values of
$K$ and types of time cut. The template of such a \PRISM model is
illustrated in Table~\ref{table:APk}.

\begin{table}[h!]
\ra{1.2}
{\scriptsize
\begin{center}
\begin{tabular}{l}
  \hline
   $\mathbf{module~} \mathit{APk}$\\
  \  $x:[0\,..\,14]~\mathbf{init}~0;$\\
  \\
  \  $[\ ]~ (x=0) \longrightarrow \Phi_{k,i,1}: (x'=1) + \ldots
 + \Phi_{k,i,14}: (x'=14);$ \\
\ \vdots\\
$\mathbf{endmodule}$
\\
\\
$\mathbf{label}~\mathit{"TermsAndConditions"} = (x=0);$\\
$\mathbf{label}~\mathit{"TopApps"} = (x=1);$\\
\vdots
\\
$\mathbf{rewards}~\mathit{"r\_TermsAndConditions"} 
\ (x=0) : 1; \
\mathbf{endrewards}$\\
$\mathbf{rewards}~\mathit{"r\_TopApps"} 
\ (x=1) : 1; \
\mathbf{endrewards}$
\\
\vdots
\\
$\mathbf{rewards}~\mathit{"r\_Steps"}
\ [\ ]\ \mathit{true} : 1;\ 
\mathbf{endrewards}$
 \\
\hline
\end{tabular}
\end{center}
}
\caption{\PRISM model for activity pattern APk}
\label{table:APk}
\end{table}

We implemented the EM algorithm in Java, applying the algorithm to
datasets with $100$ iterations maximum and $200$ restarts maximum.
Running the EM algorithm takes about 119s for $K=2$, 162s for $K=3$,
and 206s for $K=4$ on a 2.8GHz Intel Xeon. Timings are obtained by
running the algorithm 90 times. The algorithm is single threaded and
runs on one core.

\subsection{Temporal properties for activity  patterns}
\label{sect:pctlprops}

We use the following types of rPCTL properties to analyse activity
patterns for different states in \textsc{PRISM}.

\begin{property}\label{prop:ProbReach}
  The probability of reaching a given state labelled by $l$ for the
  first time from the initial state within $N$ time steps: $\TP_{=?} [\,!\, l\, \TU^{\leq N}\, l\,]$.
\end{property}

\begin{property}\label{prop:StateReward}
  The expected number of visits to a given state labeled by $l$ from
  the initial state within $N$ time steps: $\TR\{"r\_l"\}_{=?}[\TC ^{\leq N}]$.
\end{property}

\begin{property}\label{prop:Steps2StateReward}
  The expected number of time steps to reach a given state from the
  initial state: $\TR\{"\mathit{r\_Steps}"\}_{=?}[\TF\ l]$.
\end{property}

\begin{property}\label{prop:ProbReachStateFromState}
  The probability of reaching for the first time a state labelled by $l_1$ from
  another state labelled by $l_2$ during the same session: 
  $$
\ra{1.2}
\begin{array}{l}
\PropFour{l_1}{l_2}\  \eqdef\\
 \mathtt{filter}(\mathtt{state}, \TP_{=?} [ (!\,l_1 \ \&\
  !\,"\UseStop") \ U^{\leq N}\, l_1\, ],\, l_2)
\end{array}
$$
\end{property}

The next property generalises
Property~\ref{prop:Steps2StateReward} by starting from a state not
necessarily the initial one:
\begin{property}\label{prop:RewardStepsStateFromState}
  The expected number of time steps to reach a state labelled by $l_1$
  from another state labelled by $l_2$:
  $$
\ra{1.2}
\begin{array}{l}
\PropFive{l_1}{l_2} \ \eqdef \\
\mathtt{filter}(\mathtt{state}, \TR\{"\mathit{r\_Steps}"\}_{=?}[\TF\ l_1], \ l_2)
\end{array}
$$
\end{property}

Properties~\ref{prop:ProbReach}, \ref{prop:StateReward}, and
\ref{prop:Steps2StateReward} are generic temporal properties used for
sketching a first image of a DTMC. They become more app-specific when
we apply them for specific states prompted by designers'
hypotheses. Properties~\ref{prop:ProbReachStateFromState} and
\ref{prop:RewardStepsStateFromState} were identified during the rPCTL
analysis stage for specific pairs of states when more in-depth
analysis was required to make sense of the activity patterns.

While we cannot give a general interpretation of the results of the
latter two properties because they depend on the state labels, we
interpret the results of first three properties above as follows:
\begin{itemize}
\item the higher the probability computed by
  Prop.~\ref{prop:ProbReach} for the $l$-labelled state, the better;
\item the higher number of visits to the $l$-labelled state as
  computed by Prop.~\ref{prop:StateReward}, the better;
\item the fewer (non-zero) number of time steps to reach the
  $l$-labelled state as computed by
  Prop.~\ref{prop:Steps2StateReward}, the better.
\end{itemize}

\section{Analysis Results}
\label{sect:analysis}

In this section we analyse different admixture bigram models for $K$
taking values in the set $\{2,3,4,5\}$, and for various time cuts of
the logged data.  We verify rPCTL temporal properties enumerated in
Sect.~\ref{sect:pctlprops} on all activity patterns and then compare
longitudinally the weightings given by $\Theta$. In this paper we show
only properties concerning the following five states: \TopApps,
\Stats, \PeriodSelector, \LastSevenDays, \UseStop, because these
states showed significant results and differences across time cuts and
temporal properties when we analysed the entire set of states and, in
the same time, the designers showed particular interest in them when
formulating hypotheses about the actual app usage.

\begin{table*}\centering
\ra{1.3}
\caption{Property~\ref{prop:ProbReach} (the probability of reaching
  a given state for the first time within $N$ steps),
  Property~\ref{prop:StateReward} (the expected number of 
  visits to a given state within $N$ steps), and
  Property~\ref{prop:Steps2StateReward} (the expected number of 
  time steps to reach a given state) checked for different states and
  time cuts, and for $N=50$ steps} 
\vspace{2ex}
\begin{tabular}{@{}|c|c|r|r|r|r|r|r|r|r|r|r|r@{}}\hline
{\bf Prop.} & {\bf Time} & \multicolumn{2}{c|}{\bf \TopApps} &  
 \multicolumn{2}{c|}{\bf \Stats} &
 \multicolumn{2}{c|}{\bf \PeriodSelector} &
 \multicolumn{2}{c|}{\bf \LastSevenDays} &
 \multicolumn{2}{c|}{\bf \UseStop} \\
\cline{3-4} \cline{5-6} \cline{7-8} \cline{9-10} \cline{11-12}
& {\bf cut} & {\bf AP1} & {\bf AP2} & {\bf AP1} & {\bf AP2}  & {\bf AP1} & {\bf
    AP2}  & {\bf AP1} & {\bf AP2}  & {\bf AP1} & {\bf AP2} \\ 
\hline \hline
\multirow{5}{*}{\begin{sideways}Property~\ref{prop:ProbReach}\end{sideways}}
& $[0,1)$ & 0.99  & 0.99  & 0.99 & 0.83  & 0.47 & 0.79 &
0.49 & 0.96 & 0.99  & 0.99 \\ 
& $[1,7)$ & 0.99 & 0.99 & 0.98 & 0.80 & 0 & 0.93 & 0 & 0.98
& 0.99  & 0.99\\
& $[7,30)$ & 0.99 & 0.99 & 0.99 & 0.64 & 0.01 & 0.94 & 0.84
& 0.96 & 0.99  & 0.99\\
& $[0,30)$ & 0.99 & 0.99 & 0.99 & 0.75 & 0.21 & 0.92 & 0.44
& 0.98 & 0.99  & 0.99\\ 
& $[30,60)$ & 0.99 & 0.99 & 0 & 0.90 & 0.73 & 0.83 & 0.56 & 0.98 & 1 &
0.99 \\
& $[60,90)$ & 1 & 0.95 & 0.96 & 0.72 & 0 & 0.94 & 0 & 0.97
& 1 & 0.99 \\
\hline
\multirow{5}{*}{\begin{sideways}Property~\ref{prop:StateReward}\end{sideways}}
& $[0,1)$ & 13.94 & 7.44 & 7.63 & 2.15 & 0.79 & 1.82 & 0.70
& 3.13 & 11.41 & 6.17 \\
& $[1,7)$ & 17.22 & 5.77 & 4.00 & 2.31 & 0 & 3.97 & 0 &
4.03 & 12.91 & 6.30 \\
& $[7,30)$ & 14.93 & 7.15 & 5.43 & 1.47 & 0.01 & 4.61 & 1.78
& 3.41 & 12.86 & 5.74 \\
& $[0,30)$ & 14.67 & 6.48 & 5.08 & 1.90 & 0.24 & 3.58 &
0.58 & 3.99 & 11.00 & 6.51 \\
& $[30,60)$ & 13.40 & 6.83 & 0 & 3.76 & 4.41 & 2.04 & 0.85 & 4.54 &
12.46 & 5.61\\
& $[60,90)$ & 17.30 & 5.83 & 2.94 & 2.60 & 0 & 3.26 & 0 &
4.43 & 13.96 & 5.63 \\ 
\hline
\multirow{5}{*}{\begin{sideways}Property~\ref{prop:Steps2StateReward}\end{sideways}}
& $[0,1)$ & 3.31 & 8.41 & 8.18 & 28.67 & 79.32 & 32.46 &
74.87 & 15.56 & 4.86 & 7.88 \\
& $[1,7)$ & 2.05 & 10.70 & 12.44 & 31.90 & $\infty$ & 19.12
& $\infty$ & 12.38 & 3.85 & 7.55 \\
& $[7,30)$ & 2.52 & 9.68 & 9.70 & 48.61 & $\infty$ & 17.78
& 26.61 & 14.58 & 3.88 & 8.44 \\
& $[0,30)$ & 3.05 & 9.73 & 11.01 & 36.03 & 209.68 & 19.94 &
87.54 & 12.19 & 4.67 & 7.43  \\
& $[30,60)$ & 4.04 & 10.34 & $\infty$ & 22.33 & 38.21 & 28.28 & 61.74 & 11.08
& 1  & 8.82\\
& $[60,90)$ & 2.02 & 15.28 & 16.53 & 39.68 & $\infty$ &
17.41 & $\infty$ & 11.56 & 3.57 & 8.90
\\ 
\hline
\end{tabular}
\label{table:props1to3K2}
\end{table*}

\subsection{rPCTL analysis for $K=2$}

We verify Prop.~\ref{prop:ProbReach},~\ref{prop:StateReward}
and~\ref{prop:Steps2StateReward} on the two activity patterns AP1 and
AP2 for six time cuts: first day $[0,1)$, first week minus the first
day $[1,7)$, the first month minus the first week $[7,30)$, the first
month $[0,30)$, the second month $[30,60)$ and the third month
$[60,90)$, and for $N$ ranging from $10$ to $150$ with step-size
$10$. The results for $N=50$ are listed in
Table~\ref{table:props1to3K2}; the conclusions drawn in this section
hold also for the other values of $N$ we considered. In the following
we interpret the results from Table~\ref{table:props1to3K2}.

Property~\ref{prop:ProbReach} computes the probability of reaching for
the first time a given state within 50 time steps. Both AP1 and AP2
have very good results for \TopApps\ and \UseStop. AP2 has better
results for \PeriodSelector\ and \LastSevenDays\ than AP1, with
\LastSevenDays\ slightly more popular than \PeriodSelector; \Stats\
also discriminates between the two activity patterns -- better results
for AP1 than for AP2, except for the time interval $[30,60)$.

Property~\ref{prop:StateReward} computes the expected number of visits
to a state within 50 time steps while
Property~\ref{prop:Steps2StateReward} returns the expected number of
time step taken before reaching a state.  We see similar results for
these two properties: AP1 has better results for \TopApps\ and \Stats\
than AP2, while AP2 has better results for \PeriodSelector\ and
\LastSevenDays\ than AP1.  The $\infty$ results indicate that the
state is unlikely to be reached, therefore we can treat such results
as zero for this property. A session is delimited by two \UseStop\
states, except the initial session which starts from the initial
state. Thus by looking at \UseStop, on average we see twice as many
sessions under AP1 than under AP2 and the average session length in
terms of time steps under AP2 is double the average session length
under AP1.

The three properties above show slightly different results for the
time interval $[30,60)$ compared to the more consistent results for
the other five time intervals. For this interval, a high number of
visits to and a relative low number of time steps to reach \Stats\ no
longer belongs to AP1, but to AP2. Also, Prop.~\ref{prop:ProbReach}
and Prop.~\ref{prop:StateReward} for \PeriodSelector\ no longer
discriminate clearly between AP1 and AP2 due to very close results. As
a consequence we analyse additional rPCTL properties (see
Table~\ref{table:propsK2interval30to60}) for the time interval
$[30,60)$ in order to gain more insight into the two activity
patterns. The results show that under AP1 it is very unlikely to go to
\PeriodSelector\ and \LastSevenDays\ from \TopApps\, and also it is
unlikely to move between \PeriodSelector\ and \LastSevenDays, while
under AP2 these behaviours are more likely. The expected numbers of
time steps to \PeriodSelector\ and \LastSevenDays\ from \TopApps\ are
lower under AP1 than in AP2, while it takes fewer steps to reach
\TopApps\ from \PeriodSelector\ or \LastSevenDays\ under AP1. Also it
takes fewer steps to move between \PeriodSelector\ and \LastSevenDays\
under AP2 than under AP1. It takes fewer time steps on average to
reach \TopApps\ from \Main\ than it takes to reach \PeriodSelector\ or
\LastSevenDays\ under AP1 than under AP2, and vice versa. All these
results tell us that the two activity patterns learnt from the time
cut $[30,60)$ are respectively similar to the two activity patterns
earned for the rest of time cuts analysed. The difference in the
behaviour around \Stats\ could be explained by a new usage behaviour
of the AppTracker around the $30^{th}$ day of usage due to a full
month worth of new statistics, leading to a spurt of more exploratory
usage of AppTracker. We note that in the third month of usage, the
time cut $[60,90)$, the results listed in
Table~\ref{table:props1to3K2} make again a clear distinction between
the two activity patterns with respect to the states \PeriodSelector\
and \LastSevenDays. We might say that in the third month the
exploratory usage of AppTracker settles down and users know exactly
what to look for and where. A finer-grained longitudinal analysis
based on one-week time cuts could reveal additional insight into the
behaviour involving \Stats\ around the $30^{th}$ day of usage.

\begin{table*}\centering
\ra{1.3}
\caption{Properties~\ref{prop:ProbReachStateFromState}
  and~\ref{prop:RewardStepsStateFromState} verified for $K=2$, time
  interval $[30,60)$.} 
\vspace{2ex}
{\scriptsize
\begin{tabular}{|l|r|r|}
\hline
{\bf Property~\ref{prop:ProbReachStateFromState}} & {\bf AP1} & {\bf AP2} \\
\hline\hline
$\PropFour{\TopApps}{\PeriodSelector}$
  & 0.26 &  0.30 \\
$\PropFour{\PeriodSelector}{\TopApps}$
&  0.008 & 0.23 \\
\hline
$\PropFour{\TopApps}{\LastSevenDays}$
 & 0.66 & 0.26\\
$\PropFour{\LastSevenDays}{\TopApps}$
&  0.006 &  0.49 \\
\hline 
$\PropFour{\PeriodSelector}{\LastSevenDays}$
&  0.08  &  0.15 \\  
$\PropFour{\LastSevenDays}{\PeriodSelector}$
&  0.02 &  0.38\\
 \hline
\end{tabular}
}
\\ \vspace{3ex}
\begin{tabular}{cc}
{\scriptsize
\begin{tabular}{|l|r|r|}
\hline
{\bf Property~\ref{prop:RewardStepsStateFromState}} & {\bf AP1} & {\bf AP2} \\
\hline\hline
$\PropFive{\TopApps}{\PeriodSelector}$
& 11.09 & 14.89\\
$\PropFive{\PeriodSelector}{\TopApps}$
& 38.30 & 31.21 \\
\hline $\PropFive{\TopApps}{\LastSevenDays}$
&	  3.62   &	  10.56 \\
$\PropFive{\LastSevenDays}{\TopApps}$
&	  61.83  &	  14.01\\
\hline $\PropFive{\LastSevenDays}{\PeriodSelector}$
	&   68.68  & 	  15.61\\
$\PropFive{\PeriodSelector}{\LastSevenDays}$
&	  36.69  &	  28.77\\
\hline
\end{tabular}
}
&
{\scriptsize
\begin{tabular}{|l|r|r|}
\hline
{\bf Property~\ref{prop:RewardStepsStateFromState}} & {\bf AP1} & {\bf AP2} \\
\hline\hline
$\PropFive{\TopApps}{\Main}$
&	  2.22 &	  9.347   \\
$\PropFive{\PeriodSelector}{\Main}$ 
&	  36.03 &	  27.28  \\
$\PropFive{\LastSevenDays}{\Main}$
 &	  59.56  &	  10.08\\
\hline 
$\PropFive{\UseStop}{\PeriodSelector}$
&	  9.23 &	  11.51\\
$\PropFive{\UseStop}{\TopApps}$
&	  1.49 &	  11.74 \\
$\PropFive{\UseStop}{\LastSevenDays}$
&	  3.61 &	  6.24  \\
 \hline
\end{tabular}
}
\end{tabular}
\label{table:propsK2interval30to60}
\end{table*}

We conclude that there are two distinct activity patterns:
\begin{itemize}
\item {\bf \OverallAP\ pattern} corresponds to viewing mainly
  \TopApps\ and \Stats, thus more higher level stats visualisations, and 
\item {\bf \TimePartitionedAP\ pattern} corresponds to viewing in
  particular \LastSevenDays\ and \PeriodSelector, and also to some
  extent viewing \TopApps\ and \Stats\ (but less than for the
  \OverallAP\ pattern), thus more in-depth stats visualisations.
\end{itemize}

This conclusion meets the developers' hypothesis about two distinct
usages of the apps.  However developers expected also to see one
pattern revolving around \TopApps\ and \Stats, one around
\PeriodSelector\, and another one around \LastSevenDays. The choice of
$K=2$ showed only 2 distinct patterns, the last two patterns
conjectured by developers being aggregated into a single one. As a
consequence, we investigate higher values for $K$ in
Sect.~\ref{sect:analysisK3}.

\subsection{$\Theta$-based Longitudinal Analysis for $K=2$}

\begin{figure*}[!t]
  \centering 
  \subfigure[\OverallAP\ activity pattern]{\includegraphics[scale=0.55]{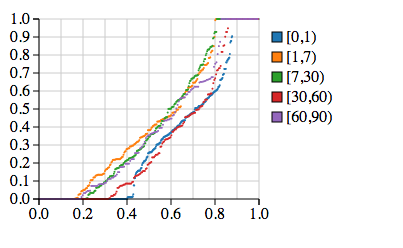}\label{fig:overallK2}}
  \subfigure[\TimePartitionedAP\ activity pattern]{\includegraphics[scale=0.55]{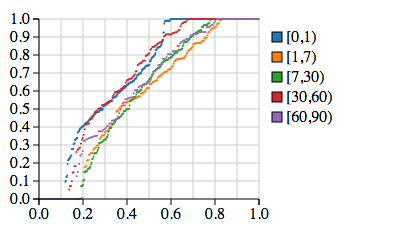}\label{fig:timepartitionedK2}}
  \caption{Longitudinal comparison of the activity pattern
    distributions $\Theta$ over the population of users for $K=2$ and
    time cuts $[0,1)$, $[1,7)$, $[7,30)$, $[30,60)$, $[60,90)$}
\label{fig:thetaK2comparison}
\end{figure*}

In addition to analysing rPCTL properties, we also compare how the
distribution $\Theta$ of the two activity patterns for the entire
population of users changes in time.  For each time cut considered for
the rPCTL analysis above and activity pattern AP2, we order
non-decreasingly the second column of $\Theta$ and re-scale its size
to the interval $[0,1]$ to represent the horizontal axis, while the
ordered $\Theta$ values are projected on the vertical axis.
Figure~\ref{fig:thetaK2comparison} shows the $\Theta$ values for AP2 for the
population of user across the first three months of usage. Since each
row of $\Theta$ sums up to 1, it is easy to picture a similar chart
for AP1. We conclude that during the first day of usage, up to 40\% of
users exhibit exclusive \TimePartitionedAP\ behaviour (probability
close to 1 on the $y$-axis) corresponding to an initial exploration of
the app with significant number of visits to \TopApps, \Stats,
\PeriodSelector, and \LastSevenDays. Also, at most 10\% of the users
exhibit exclusive \OverallAP\ behaviour maybe because they feel less
adventurous in exploring the app, preferring mostly the first menu
option of looking at \TopApps\ and subsequently at \Stats. We note
that the distributions of the two activity patterns in the population
of users are similar for the time intervals $[0,1)$ and $[30,60)$ -- 
probably because more users exhibit a more exploratory behaviour
during these times (new types of usage statistics become available
after one month of usage). At the same time, the plots for the time
intervals $[1,7)$, $[7,30)$, and $[60,90)$ are also similar, and we
think that they correspond to a settled (or routine) usage behaviour.

\subsection{rPCTL Analysis for $K= 3$}
\label{sect:analysisK3}

Let us consider the admixture model inferred for $K=3$.  We verify
Props.~\ref{prop:ProbReach},~\ref{prop:StateReward}, and
~\ref{prop:Steps2StateReward} on the time interval $[0,30)$ (first
month of usage) and $N=50$, and list the results in
Table~\ref{table:propsK3interval0to30}. Based on these results we
conclude that: 
\begin{itemize}
\item AP1 is an \OverallAP\ pattern because \TopApps\ and \Stats\ have
  best results for all three properties; \PeriodSelector\ and
  \LastSevenDays\ are absent. 
\item AP2 is an \OverallAP\ pattern 'weaker' than AP1 because
  \TopApps\ has poorer results, and better results than \Stats\ and
  \LastSevenDays; \PeriodSelector\ is absent.
\item AP3 is a \TimePartitionedAP\ pattern because \PeriodSelector\
  has the best results, followed closely by \TopApps\ and
  \LastSevenDays. 
\end{itemize}
As for $K=2$, the sessions for the \OverallAP\ pattern are twice as
short and twice more frequent than for the \TimePartitionedAP\
pattern.

\subsection{rPCTL Analysis for $K= 4$}
\label{sect:analysisK4}

We now consider the admixture model inferred for $K=4$. 
We verify Props.~\ref{prop:ProbReach},~\ref{prop:StateReward},
and~\ref{prop:Steps2StateReward} on the time interval $[0,30)$ (first
month of usage) and for $N=50$, and list the results in
Table~\ref{table:propsK4interval0to30}.  Based on these results we
conclude that:
\begin{itemize}
\item AP1 is mainly a \TopApps\ \textsf{Viewing} activity pattern
  because it has the best results for \TopApps, compared to \Stats,
  \PeriodSelector, and \LastSevenDays\ which score very weak results.
\item AP2 is a \Stats\ -- \TopApps\ \textsf{Viewing} activity pattern,
  with very weak results from \LastSevenDays, \PeriodSelector\ is
  absent.
\item AP3 is a \TimePartitionedAP\ pattern with dominant
  \LastSevenDays\ followed closely by \TopApps\ and \PeriodSelector;
  \Stats\ is absent. 
\item AP4 is mainly a \TopApps\ \textsf{Viewing} pattern \ because
  \TopApps\ has the best results, while all other states need on
  average an infinite number of time steps to be reached. The fact
  that it takes on average an infinite number of time steps to reach
  the end of a session (i.e., the state \UseStop), pushed us to
  analyse this pattern with other temporal properties and for other
  states. As a consequence we saw that \UsageBarChartTopApps\ has
  similar properties as \TopApps, meaning that this pattern
  corresponds to repeatedly switching between \TopApps\ and
  \UsageBarChartTopApps.
\end{itemize}

By verifying the Props.~\ref{prop:ProbReach},~\ref{prop:StateReward},
and~\ref{prop:Steps2StateReward} also for the state \UseStop\ we
observe twice as many sessions for AP1 than for AP2 and AP3, only a
couple of sessions on average for AP4, fewer time steps per session
for AP1 than for AP2 and AP3.

\subsection{rPCTL Analysis for $K= 5$}
\label{sect:analysisK5}

Let us consider the admixture model inferred for $K=5$. We analyse
Props.~\ref{prop:ProbReach},~\ref{prop:StateReward},
and~\ref{prop:Steps2StateReward} on the time interval $[0,30)$ (first
month of usage) and for $N=50$, and list the results in
Table~\ref{table:propsK5interval0to30}.  Based on these results we
conclude that:
\begin{itemize}
\item AP1 is mainly a \TopApps\ \textsf{Viewing} activity pattern with
  very long sessions in average, thus pointing towards the pattern
  corresponding to the loop between \TopApps\ and
  \UsageBarChartTopApps\ (as confirmed by further analysis of the
  Props.~\ref{prop:ProbReach},~\ref{prop:StateReward},
  and~\ref{prop:Steps2StateReward} for the \UsageBarChartTopApps\
  state).
\item AP2 is a \TimePartitionedAP\ pattern with dominant
  \PeriodSelector\ and slightly less \LastSevenDays, and almost no
  \TopApps\ or \Stats.
\item AP3 is an \OverallAP\ pattern weaker than AP4 and AP5 because
  \TopApps\ has poorer results, and better results than \Stats\ and
  \LastSevenDays; \PeriodSelector\ is absent.
\item AP4 is an \OverallAP\ pattern because \TopApps\ has the best
  results, \Stats\ comes in second. \PeriodSelector\ and
  \LastSevenDays\ are absent.
\item AP5 is an \OverallAP\ pattern weaker than AP4 with \Stats\
  scoring slightly better results than \TopApps, and \LastSevenDays\
  scores poorer results, while \PeriodSelector\ is absent.
\end{itemize}

\begin{figure*}[!t]
  \centering 
  \includegraphics[scale=0.4]{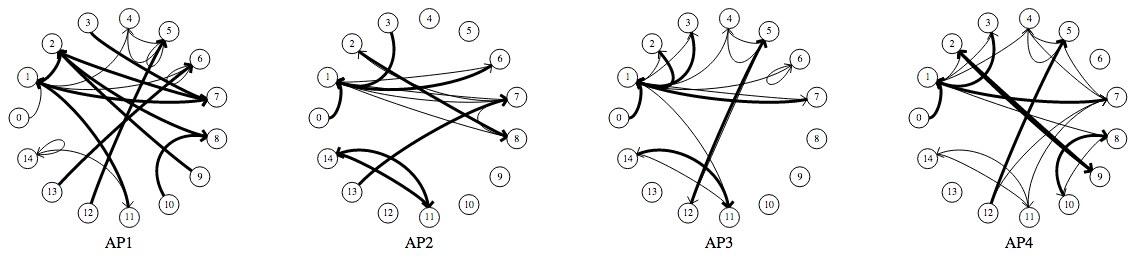}
  \caption{The DTMCs of the activity patterns for $K=4$ where states
    are enumerated from $0$ to $14$}
\label{fig:patternsK4}
\end{figure*}

We do not show the probability transition matrix for the activity
patterns analysed in this section for the sake of simplicity. The
properties investigated here create an overall image of the likelihood
of correlations between what we identified as most relevant five
states. Figure~\ref{fig:patternsK4} illustrates the state-transition
diagrams of all $K=4$ activity patterns with thickness of the
transitions corresponding to ranges of transition probabilities: the
thicker the line, the higher the probability of that transition, while
transitions with very small probabilities are not shown. Such graphs
offer a high-level characterisation of patterns, but we note that a
high probability from one state $s$ to a state $s'$ in such a graph
does not mean that the transition is very likely within the execution
of the DTMC: if the probability of reaching $s$ is very low, then the
transition from $s$ to $s'$ will have a low probability to take place
during execution. However, when considering the characteristics of
each pattern as discovered through rPCTL analysis, such graphs aid in
pattern identification and understanding.  In particular AP4 stands
out with a very thick transition between states \TopApps\ (state id 2)
and \UsageBarChartTopApps\ (state id 9).

\begin{table*}\centering
\ra{1.3}
\caption{Model checking Prop.~\ref{prop:ProbReach} (the probability
  of reaching 
  a given state for the first time within $50$ time steps),
  Prop.~\ref{prop:StateReward} (the expected number of 
  visiting a given state within $50$ steps), and
  Prop.~\ref{prop:Steps2StateReward} (the expected number of 
  time steps to reach a given state) checked for five states,
  time cut $[0,30)$, and $K=3$} 
\vspace{2ex}
\begin{tabular}{|c|rrr|rrr|rrr|rrr|rrr|}
\hline
{\bf Prop.} & 
 \multicolumn{3}{c|}{\bf \TopApps} & 
 \multicolumn{3}{c|}{\bf \Stats} & 
 \multicolumn{3}{c|}{\bf \PeriodSelector} &
 \multicolumn{3}{c|}{\bf \LastSevenDays} 
 & \multicolumn{3}{c|}{\bf \UseStop} \\
\cline{2-4} \cline{5-7} \cline{8-10} \cline{11-13} \cline{14-16}
& AP1 & AP2 & AP3 & AP1 & AP2 & AP3 & AP1 & AP2 & AP3 & AP1 & AP2 &
AP3 & AP1 & AP2 & AP3 \\ \hline\hline
\ref{prop:ProbReach} & 0.99 & 0.99 & 0.93 & 0.99 & 0.91 & 0.39 & 0 & 0
& 0.97 & 0 & 0.91 & 0.96 & 0.99 & 0.99 & 0.99\\
\ref{prop:StateReward} & 15.45 & 6.94 & 6.74 & 5.56 & 2.57 & 0.99 & 0
& 0 & 8.46 & 0 & 2.32 & 3.98 & 13.65 & 8.20 & 6.00 \\
\ref{prop:Steps2StateReward}  & 2.45 & 7.35 & 17.43 & 9.33 & 22.01 &
98.96 & $\infty$ & $\infty$ & 11.74 & $\infty$ & 20.99 & 12.55 & 3.78
& 5.86 & 8.08 \\
\hline
\end{tabular}
\label{table:propsK3interval0to30}
\end{table*}

\begin{table*}\centering
\ra{1.3}
\caption{Model checking Prop.~\ref{prop:ProbReach} (the probability of reaching
  a given state for the first time within $50$ time steps),
  Prop.~\ref{prop:StateReward} (the expected number of 
  visiting a given state within $50$ steps), and
  Prop.~\ref{prop:Steps2StateReward} (the expected number of 
  time steps to reach a given state) checked for five states,
  time cut $[0,30)$, and $K=4$} 
\vspace{2ex}
\begin{tabular}{|c|rrrr|rrrr|}
\hline
{\bf Prop.} & 
 \multicolumn{4}{c|}{\bf \TopApps} & 
 \multicolumn{4}{c|}{\bf \Stats} \\
\cline{2-5} \cline{6-9} 
& AP1 & AP2 & AP3 & AP4 & AP1 & AP2 & AP3 & AP4  \\ \hline\hline
\ref{prop:ProbReach} & 0.99 & 0.99& 0.99& 0.95& 0.04& 0.99& 0& 0.06 \\
\ref{prop:StateReward} & 15.79& 9.93& 4.64& 17.84& 0.03& 10.85& 0&
0.19 \\
\ref{prop:Steps2StateReward} & 2.64&
4.36& $\infty$& 
$\infty$& $\infty$& $\infty$& 26.51& $\infty$\\
\hline
\end{tabular}
\\
\vspace{3ex}
\begin{tabular}{|c|rrrr|rrrr|rrrr|}
\hline
{\bf Prop.} & 
 \multicolumn{4}{c|}{\bf \PeriodSelector} &
 \multicolumn{4}{c|}{\bf \LastSevenDays} &
 \multicolumn{4}{c|}{\bf \UseStop} \\
\cline{2-5} \cline{6-9} \cline{10-13}  
& AP1 & AP2 & AP3 & AP4 & AP1 & AP2 & AP3 & AP4 & AP1 & AP2 & AP3 &
AP4  \\ \hline\hline
\ref{prop:ProbReach} & 0.06& 0& 0.85& 0.57& 0.02& 0.13& 0.99& 0.58 &
0.99 & 0.99 & 0.99 & 0.66\\
\ref{prop:StateReward} & 0.09& 0 & 2.97& 3.03& 0.02& 0.14& 5.70& 1.34
& 15.59 & 7.46 & 7.39 & 2.14 \\
\ref{prop:Steps2StateReward} & $\infty$& 378.76& 8.46& $\infty$& 3.49&
7.30& 6.54& 0 & 3.49 & 7.30 & 6.54 & $\infty$\\ 
\hline
\end{tabular}
\label{table:propsK4interval0to30}
\end{table*}

\begin{table*}\centering
\ra{1.3}
\caption{Model checking Prop.~\ref{prop:ProbReach} (the probability of reaching
  a given state for the first time within $50$ time steps),
  Prop.~\ref{prop:StateReward} (the expected number of 
  visiting a given state within $50$ steps), and
  Prop.~\ref{prop:Steps2StateReward} (the expected number of 
  time steps to reach a given state) checked for five states,
  time cut $[0,30)$, $K=5$} 
\vspace{2ex}
\begin{tabular}{|c|rrrrr|rrrrr|}
  \hline
  {\bf Prop.} & 
  \multicolumn{5}{c|}{\bf \TopApps} & 
  \multicolumn{5}{c|}{\bf \Stats} \\
  \cline{2-6} \cline{7-11} 
  & AP1 & AP2 & AP3 & AP4 & AP5 & AP1 & AP2 & AP3 & AP4 & AP5  \\ 
\hline\hline
  \ref{prop:ProbReach} & 0.99 & 0.06 & 0.99 & 0.99 & 0.99 & 0.19 & 0.07
  & 0.71 & 0.86 & 0.99  \\ 
  \ref{prop:StateReward} & 21.65 & 0.07 & 6.61 & 17.40 & 5.37 & 0.54 &
  0.10 & 1.40 & 1.87 & 7.56  \\
  \ref{prop:Steps2StateReward}  & 5.06 & 721 & 7.91 & 2.09 & 7.13 &
  211.08 & 600 & 40.28 & 26.42 & 6.43 \\ 
  \hline
\end{tabular}
\\
\vspace{3ex}
\begin{tabular}{|c|rrrrr|rrrrr|rrrrr|}
  \hline
  {\bf Prop.} & 
  \multicolumn{5}{c|}{\bf \PeriodSelector} &
  \multicolumn{5}{c|}{\bf \LastSevenDays} 
  & \multicolumn{5}{c|}{\bf \UseStop} \\
  \cline{2-6} \cline{7-11} \cline{12-16}
  & AP1 & AP2 & AP3 & AP4 & AP5 & AP1 & AP2 & AP3 & AP4 & AP5 & AP1 &
  AP2 & AP3 & AP4 & AP5  \\ \hline\hline
  \ref{prop:ProbReach} & 0.15 & 0.99 & 0 & 0 & 0 & 0.34 & 0.99 & 0.92
  & 0 & 0.54 & 0.35 & 0.99 & 0.99 & 0.99 & 0.99 \\ 
  \ref{prop:StateReward} & 0.46 & 11.47 & 0 & 0 & 0 & 0.50 & 6.05 &
  2.50 & 0 & 0.76 & 0.49 & 11.15 & 9.81 & 15.82 & 9.23 \\
  \ref{prop:Steps2StateReward}  & 1059.91 & 5.44 & $\infty$ &
  $\infty$ & $\infty$ & 369 & 7.13 & 19.64 & $\infty$ & 64.41 & 340 &
  4.46 & 4.95 & 3.20 & 5.72 \\ 
  \hline
\end{tabular}
\label{table:propsK5interval0to30}
\end{table*}

\subsection{$\Theta$-based analysis for the time cut $[0,30)$}

\begin{figure*}[!t]
  \centering 
  \subfigure[$K=2$]{\includegraphics[scale=0.55]{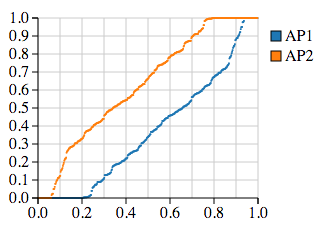}\label{fig:thetaK2cut0to30}}
  \quad
  \subfigure[$K=3$]{\includegraphics[scale=0.55]{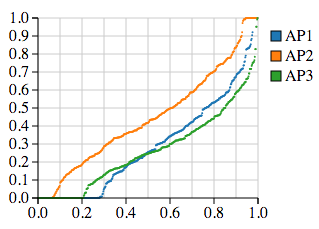}\label{fig:thetaK3cut0to30}}
  \\
  \subfigure[$K=4$]{\includegraphics[scale=0.55]{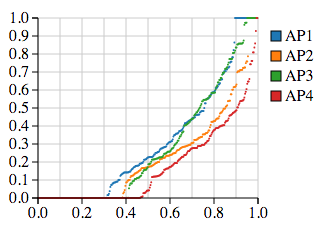}\label{fig:thetaK4cut0to30}} 
  \quad
  \subfigure[$K=5$]{\includegraphics[scale=0.55]{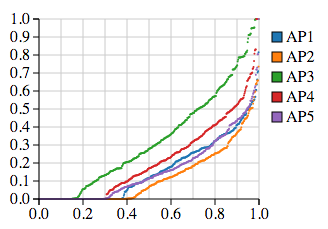}\label{fig:thetaK5cut0to30}}
  \caption{Pattern distributions for $K=2$ (\OverallAP,
    \TimePartitionedAP), $K=3$ (\OverallAP, weak \OverallAP,
    \TimePartitionedAP), $K=4$ (mainly \TopApps\ Viewing, equally
    \Stats\ and \TopApps\ Viewing, \TimePartitionedAP\ with no \Stats,
    exclusive \TopApps\ and \UsageBarChartTopApps), $K=5$ (exclusive
    \TopApps\ and \UsageBarChartTopApps\ Viewing, \TimePartitionedAP,
    weakest \OverallAP, \OverallAP, weak \OverallAP), time cut
    $[0,30)$.}
\label{fig:thetaKcomparison}
\end{figure*}

In Fig.~\ref{fig:thetaKcomparison} we plot the weightings of all 485
users in the time cut $[0,30)$ for each activity patterns for
$K\in\{2,5\}$.  Figure~\ref{fig:thetaK2cut0to30} tells us that for
$K=2$ the \TimePartitionedAP\ has higher weightings across the user
population with almost 25\% of the users using the app exclusively
like this, hence either exploring the app or genuinely interested in
in-depth usage statistics.  Figure~\ref{fig:thetaK3cut0to30} tells us
that almost 10\% of the users are exclusively interested in \TopApps,
\Stats\ and \LastSevenDays\, but not \PeriodSelector; this behaviour
is the most popular among users. From Fig.~\ref{fig:thetaK4cut0to30}
we see that almost 50\% of the users do not behave according to AP4 --
switching repeatedly between \TopApps\ and \UsageBarChartTopApps. Note
that for $K=3$, $K=4$, and $K=5$ no pattern stands out as very
different than the others.

\section{Informing App Redesign}
\label{sect:redesign}

The results of our analysis offer several insights into the usage of
AppTracker, which can be drawn upon by developers in the design of
future versions.

Our analysis initially uncovered two activity patterns for
AppTracker. These patterns can largely be characterised by the type of
app usage data the user is examining: either \OverallAP\ -- more
higher level usage statistics visualisations for the entire recorded
period, or \TimePartitionedAP\ -- more in-depth usage statistics
visualisation for specific periods of interest. We have not found one
usage pattern to be significantly dominant over the other. For the
majority of users, usage is fairly evenly distributed between the two
patterns, suggesting that a revised version of AppTracker should
continue to support both rather than focusing on only one.

The two patterns identified in the admixture model with $K=2$
correspond quite closely to options presented on AppTracker's main
menu (see Fig.~\ref{fig:menu}), which is the initial page shown when
AppTracker launches. \OverallAP\ shows more usage of \TopApps\ and
\Stats, which are interface screens accessed through the {\em Overall
  Usage} menu item. \TimePartitionedAP\ sees higher probabilities for
reaching \PeriodSelector\ and \LastSevenDays, which are accessed
through {\em Select by Period} and {\em Last 7 Days}, but also some
usage of \TopApps\ and \Stats. Our results indicate that usage
sessions corresponding to \OverallAP\ are generally shorter. This
means that during these sessions, users are performing fewer actions
between launching AppTracker and exiting back to the device's home
screen. The diverse patterns suggest that, in a future version of
AppTracker, if developers wanted to keep the two major styles of usage
separated between different screens, they could explicitly design for
these {\em glancing}-like short interactions in {\em Overall Usage}
and longer interactions in a new {\em Select by Period} screen along
with the initial {\em Last 7 Days} screen. Also more filtering and
querying tools could be added to {\em Select by Period}.


In noting that activity patterns divide between main menu options in
this way, we might be concerned that users are simply following the
suggested paths as defined by the interface.  Have our processes
uncovered the inherent usage styles of AppTracker that users favour,
and revealed that the initial menu design was sensible, matching these
inherent styles well? Or is it the case that the menu acts as a prompt
and users are simply following these suggestions given by the
interface? If this latter explanation was correct, then it could be
argued that our analysis has merely recovered AppTracker's menu
structure.  We therefore probed further, running further admixture
bigram models for $K=3$, $K=4$ and $K=5$. For the case of $K=3$, if
the analysis was merely mirroring the menu structure, we might expect
to see one pattern centered around each of the first three menu items.
Although we see a pattern AP2 centered around \TopApps, \Stats, and
\LastSevenDays\ but no \PeriodSelector, we do not see a pattern
centered around \PeriodSelector\ and not including \LastSevenDays. For
$K=4$ we find these two views together in a pattern, with
\LastSevenDays\ slightly more popular than \PeriodSelector.  This
combination also occurred for $K=2$ and $K=5$.

For $K=4$, we see a distinct new cluster of activity showing users
repeatedly switching between \TopApps\ and
\UsageBarChartTopApps. \TopApps\ is an ordered list of the user's most
used apps; selecting an item from this list opens
\UsageBarChartTopApps, a bar chart showing daily minutes of use of
this app. This persistent switching suggests a more investigatory
behaviour, which is more likely to be associated with the
\TimePartitionedAP. Yet this behaviour is occurring under the {\em
  Overall Usage} menu item which we hypothesised and then identified
as being associated with more glancing-like behaviour. This suggests
that our results are providing more nuanced findings than simple
uncovering of existing menu structure.  Also, if developers wanted to
separate the two types of usage between different menu items even
more, they could move the \TopApps\ -- \UsageBarChartTopApps\ loop
from {\em Overall Usage} to {\em Select by Period}).

Discovering {\em glancing} activity patterns has significant benefits
in app redesign. Since the release of the iOS 8 SDK in 2014, Apple has
allowed the development of 'Today widgets'. These are extensions to
apps comprising of small visual displays and limited functionality
that appear in the device's Notification Centre, accessible by swiping
down from the top of the screen. On the subject of Today widgets,
Apple's Human Interface Guidelines
(HIG)~\footnote{\url{https://developer.apple.com/library/ios/documentation/UserExperience/Conceptual/MobileHIG/AppExtensions.html}}
state that {\em ``it's best when your Today widget displays the right
  amount of information and limits interactivity''}, encouraging
developers to {\em ``keep user interactions limited and streamlined''}
and explaining that {\em ``Because your Today widget provides a
  narrowly focused experience, it can work well to direct people to
  your app for more information or functionality.''}  Beyond these
pieces of advice, however, developers might struggle to decide which
pieces of their app's contents would best suit inclusion in a Today
widget. Few conventions have built up on this in the limited time the
SDK has been available, and most developers would have to rely on
their own judgement to select appropriate content from their app to
populate this view. In our analysis, we have explicitly uncovered the
specific screens that people look at when they are undertaking short
sessions of glancing-type behaviour, i.e. the typical glancing
patterns for AppTracker -- the \OverallAP\ pattern and the
\TopApps-centered patterns. In identifying such activity patterns, our
methods provide a more principled method of selecting content
appropriate for an app extension such as a Today widget.

We conclude that our approach presented in this paper is one
additional way of influencing the iterative process of redesign while
collecting more data based on updated release and analysing them.

\section{Related Work}
\label{sect:relatedwork}
 
Logging software is also frequently used to understand program
behaviours, and typically to aid program comprehension -- building an
understanding of how the program executes~\cite{ProgComp95}. There are
various techniques that use logs of running software, such as
visualising logs (e.g.~\cite{ExploreViz}) and capture and replay
(e.g.~\cite{Mickens:2010},~\cite{Gomez:2013}). The motivations for
doing this is often failure analysis, evaluating performance, and to
better understand the system behaviour (as it executes). In contrast,
we are interested in what ways users are interacting with the
software, and we are doing so by analysing logs captured during the
use of the software. The difference here is important. In the case of
program comprehension, log analysis is used to  understand better what
is going on within the code and how the artifact is engineered (in
order to be better prepared for improvements and maintenance). In the
second case however, analysing usage logs is done to learn about
distinct uses of the software to inform improvements on the
higher-level design. There is however an interest from software
engineer practitioners to learn about the use of an app through data
science~\cite{Begel:2014}.
 
The approach of~\cite{GhezziPST14} also employs usage logs as a
resource for improvement on design, and applies temporal logical
analysis. A key difference is that their models are based on
relatively static user attributes (e.g. city location of user) rather
than on inferred behaviours.  Their approach assumes within-class use
to be homogeneous, but our research demonstrates within-class
variation. We take a different approach to inference. We have found
the common representations of context -- such as location, operating
system, or time of day -- to be poor predictors of mobile application
use in a first instance. For this reason we construct user models
based on the log information alone, without any ad-hoc specification
of static user attributes. By letting the data speak for itself, we
hope to uncover interesting activity patterns and meaningful
representations of users.
 
We first introduced the concept of representing the behaviour of users
through a weighted mixture over data generating
distributions~\cite{HiggsMGC13}, refining the concept substantially
in~\cite{AndreiCHG14} where we defined activity patterns for an
individual user as user meta models (DTMCs), with respect to a
population of users.  We then inferred behaviours for individual user
activity from large scale logged usage data for a mobile game app and
analysed them using probabilistic temporal properties (without
rewards). This current work builds on that approach, but differs
substantially in that here our goal is redesign in the context of a
different app. To this end we investigate a range of values for the
number $K$ of activity patterns and completely different temporal
properties (e.g. using rewards) for longitudinal
analysis. In~\cite{AndreiCHG14} we analysed individual user models,
whereas in this paper we analysed the whole population of users as we
compared the distributions of activity patterns across the user
population longitudinally for a fixed value $K$ as well as for three
different values for $K$.

\section{Discussion}
\label{sect:discussion} 

In this section we shift emphasis from analysis and redesign of the
AppTracker app to more general discussion of this type of analysis,
offering suggestions and guidelines for how it can be used with other
apps, and how analysts and developers may use it in their work
together.

At present, the kind of approach we describe here is likely to involve
collaboration between developers familiar with app development and
instrumentation, and analysts familiar with statistics and formal
modelling. With this in mind, we note some methodological issues that
those considering such joint work may wish to consider.

Firstly, our approach gains from having significant volumes of log
data to work on, in order to allow the statistical methods to be
applied more reliably. However, the volume of log data to analyse is
not relevant for the probabilistic model checking, only the number of
higher level states for analysis selected from the raw data determines
the complexity of the model checking problem. Our case study involved
a vocabulary of 15 states, which was comfortably manageable, but we
note that the combinatorial growth of the complexity of model checking
argues against very large vocabularies.

Our second point follows on from the first: the issue of {\em what to
  log} is not trivial. Clearly, it is possible to instrument an
app with logging in many different ways. This is typically decided
during development, and the resulting logs -- both between apps but
also within apps -- are therefore diverse. Unless it is strictly
decided precisely what to log, the collection of log entries can
include anything from a button press, to the change of WiFi signal of
the device, and so on. For the method of analysis as used in this
paper, one needs to go from these logs to transitions between
states. One therefore needs to identify what these higher level states
are, and how they correlate with the logs, as the choice of states is
the choice of the lens through which one interprets the use of the
app. In~\cite{GhezziPST14} states were related to web page views
(similar to views in AppTracker, to some extent). In contrast,
in~\cite{AndreiCHG14}, the states were related to specific game
actions in the game.  The two perspectives (and associated atomic
propositions) are different in nature, and they each capture something
different about what users do with the system under investigation.

In the case of AppTracker, we decided to use states corresponding to
individual screens possible to transition to within the app. This
highlights the rather simple nature of AppTracker -- it is essentially
a browser of information. In contrast, a game such as Angry Birds
allows the user to perform a much more complex set of actions. Even
after pruning the logs to include only user actions (rather than lower
level device events), one still needs to decide how to model these
actions as a state space. In the end, the chosen state space will
ultimately influence what activity patterns become visible. We also
note the need for care over the choice of what to log, the possibility
of revision of that choice in the light of ongoing analysis, and the
potential cost of reconciliation or integration of log data from
different logging regimes.  We therefore suggest that discussion and
preliminary analysis be done early in the development process, so that
the decisions about what to log and what the state space should be are
made by developers and analysts jointly in a well-informed way.

Thirdly, the value of $K$ is clearly key to analysis, but what is the
most appropriate value for $K$? While we could use model selection or
non-parametric methods to infer $K$, there might be reasons to fix a
value of $K$ based in domain knowledge and/or developer's knowledge of
the app. We suggest that the analyst should expect to work in an
incremental and exploratory way, using results from ongoing analysis
as well as the developers' knowledge of the app and the application
domain. In the case study, we noted that AppTracker's developers
expected that two or three particular activity patterns would be
uncovered by the analysis. Two major different patterns were in fact
found -- one of which subsumed the developers' expected patterns. The
findings from $K=2$ were useful in themselves, but also led to
analysing higher values of $K$ and thus to findings that required
developer knowledge for full interpretation.

Lastly, we note the complementarity of our methods based on inferred
behaviours to methods using attributes selected {\em a priori} (such
as~\cite{GhezziPST14}, discussed in Sect.~\ref{sect:relatedwork}), and
also to more everyday analysis such as counting how often a particular
UI state was reached (e.g. via simple SQL queries over the log
data). In our case study, we often used SQL queries and JavaScript
visualisations to explore and prepare for more complex temporal
analysis, with developers and analysts working together via those
queries to frame more specialised analytic work -- which in turn often
led on to further SQL and JavaScript work. Similarly, we suggest that
attributes selected {\em a priori} can be used to express domain
knowledge in ways that both generate useful findings and lead on to
other questions and analyses. We propose that the presented temporal
logical analysis should be seen as adding to the toolkit available to
developers, analysts and researchers, rather than competing with other
methods.

\section{Conclusions}
\label{sect:conclusions}

We have outlined our approach to informing redesign based on
probabilistic formal analysis of actual app usage.  Our approach is a
combination of bottom up statistical inference from user traces, and
top down probabilistic temporal logic analysis of inferred models.  We
have illustrated this via a mobile app, and discussed how the results of
this analysis inform app redesign that is grounded in existing patterns
of behaviour.

Future work lies in two directions. First, we have developed and
employed a prototype web-based environment for creating the time cuts,
preparing data for and running the EM algorithm, and for generating
\PRISM models from the EM outputs and visualisations. Manual
interventions are still required at various stages, particularly when
analysing \PRISM properties and generating visualisations, and ongoing
work focuses on developing an analysis framework that encompasses all
the stages within one easily adaptable web based environment, easing
and speeding up the collaborative work of analysis.  Second, our
choice of bigram admixture models is based on the work
of~\cite{GirolamiK04} in modelling web-browsing activity across
populations of individuals.  What type of probabilistic model would
help us investigate possible causes for a user to transition from one
type of behaviour to another one? Can we determine the context in
which they are likely to make that transition?  Future work involves
Hierarchical Hidden Markov models~\cite{Murphy02dynamicbayesian},
where the first abstract level in the hierarchy consists of contextual
features and then the activity patterns learned for each feature.

\section{Acknowledgments}

This research is supported by the EPSRC Programme Grant {\em A
  Population Approach to Ubicomp System Design} (EP/J007617/1).  The
authors thank all members of the project.


\bibliographystyle{abbrv}

\begin{thebibliography}{10}

\bibitem{AndreiCHG14}
O.~Andrei, M.~Calder, M.~Higgs, and M.~Girolami.
\newblock {Probabilistic Model Checking of DTMC Models of User Activity
  Patterns}.
\newblock In G.~Norman and W.~H. Sanders, editors, {\em {Proc. of QEST'14}},
  volume 8657 of {\em {Lecture Notes in Computer Science}}, pages 138--153.
  Springer, 2014.

\bibitem{BaierKatoen-MCbook}
C.~Baier and J.-P. Katoen.
\newblock {\em {Principles of Model Checking}}.
\newblock {The MIT Press}, 2008.

\bibitem{Begel:2014}
A.~Begel and T.~Zimmermann.
\newblock {Analyze This! 145 Questions for Data Scientists in Software
  Engineering}.
\newblock In {\em {Proc. of ICSE14}}, pages 12--23, New York, NY, USA, 2014.
  ACM.

\bibitem{BellCFHMRRS13}
M.~Bell, M.~Chalmers, L.~Fontaine, M.~Higgs, A.~Morrison, J.~Rooksby, M.~Rost,
  and S.~Sherwood.
\newblock {Experiences in Logging Everyday App Use}.
\newblock In {\em {Proc. of Digital Economy'13}}. ACM, 2013.

\bibitem{Demp1977}
A.~P. Dempster, N.~M. Laird, and D.~B. Rubin.
\newblock {Maximum Likelihood from Incomplete Data via the EM Algorithm}.
\newblock {\em {Journal of the Royal Statistical Society. Series B
  (Methodological)}}, 39(1):pp. 1--38, 1977.

\bibitem{ExploreViz}
F.~Fittkau, J.~Waller, C.~Wulf, and W.~Hasselbring.
\newblock {Live trace visualization for comprehending large software
  landscapes: The ExplorViz approach}.
\newblock In {\em {Proc. of VISSOFT'13}}, pages 1--4, 2013.

\bibitem{GhezziPST14}
C.~Ghezzi, M.~Pezz{\`e}, M.~Sama, and G.~Tamburrelli.
\newblock {Mining Behavior Models from User-Intensive Web Applications}.
\newblock In P.~Jalote, L.~C. Briand, and A.~van~der Hoek, editors, {\em {Proc.
  of ICSE'14}}, pages 277--287. {ACM}, 2014.

\bibitem{GirolamiK04}
M.~Girolami and A.~Kaban.
\newblock {Simplicial Mixtures of Markov Chains: Distributed Modelling of
  Dynamic User Profiles}.
\newblock In S.~Thrun, L.~Saul, and B.~{Sch\"{o}lkopf}, editors, {\em {Advances
  in Neural Information Processing Systems 16}}. MIT Press, Cambridge, MA,
  2004.

\bibitem{Gomez:2013}
L.~Gomez, I.~Neamtiu, T.~Azim, and T.~Millstein.
\newblock {RERAN: Timing- and Touch-sensitive Record and Replay for Android}.
\newblock In {\em {Proc. of ICSE'13}}, ICSE '13, pages 72--81, Piscataway, NJ,
  USA, 2013. IEEE Press.

\bibitem{Hall:2009}
M.~Hall, M.~Bell, A.~Morrison, S.~Reeves, S.~Sherwood, and M.~Chalmers.
\newblock {Adapting ubicomp software and its evaluation}.
\newblock In {\em {Proc. of EICS'09}}, pages 143--148, New York, NY, USA, 2009.
  ACM.

\bibitem{HiggsMGC13}
M.~Higgs, A.~Morrison, M.~Girolami, and M.~Chalmers.
\newblock {Analysing User Behaviour Through Dynamic Population Models}.
\newblock In {\em {Proc. of CHI'13, Extended Abstracts on Human Factors in
  Computing Systems}}, {CHI EA'13}, pages 271--276. {ACM}, 2013.

\bibitem{KwiatkowskaNP07}
M.~Z. Kwiatkowska, G.~Norman, and D.~Parker.
\newblock {Stochastic Model Checking}.
\newblock In M.~Bernardo and J.~Hillston, editors, {\em {SFM}}, volume 4486 of
  {\em {LNCS}}, pages 220--270. Springer, 2007.

\bibitem{KwiatkowskaNP11}
M.~Z. Kwiatkowska, G.~Norman, and D.~Parker.
\newblock {PRISM 4.0: Verification of Probabilistic Real-Time Systems}.
\newblock In {\em {Proc. of CAV'11}}, volume 6806 of {\em {LNCS}}, pages
  585--591. Springer, 2011.

\bibitem{Mickens:2010}
J.~Mickens, J.~Elson, and J.~Howell.
\newblock {Mugshot: Deterministic Capture and Replay for Javascript
  Applications}.
\newblock In {\em {Proc. of the 7th USENIX Conference on Networked Systems
  Design and Implementation}}, NSDI'10, pages 11--11. USENIX Association, 2010.

\bibitem{Murphy02dynamicbayesian}
K.~P. Murphy.
\newblock {\em {Dynamic Bayesian Networks: Representation, Inference and
  Learning}}.
\newblock PhD thesis, {University of California, Berkley}, 2002.

\bibitem{StollerBSGHSZ11}
S.~D. Stoller, E.~Bartocci, J.~Seyster, R.~Grosu, K.~Havelund, S.~A. Smolka,
  and E.~Zadok.
\newblock {Runtime Verification with State Estimation}.
\newblock In {\em {Proc. of RV'11}}, volume 7186 of {\em {LNCS}}, pages
  193--207. Springer, 2011.

\bibitem{ProgComp95}
A.~von Mayrhauser and A.~Vans.
\newblock {Program comprehension during software maintenance and evolution}.
\newblock {\em Computer}, 28(8):44--55, 1995.

\end{thebibliography}

\end{document}